\documentclass[11pt,twoside]{article}
\usepackage{amsmath}
\usepackage{amssymb}
\usepackage{graphicx}
\usepackage[english]{babel}
\usepackage{ihep}

\begin{document}
\newcommand{\abs}[1]{\lvert#1\rvert}
\title{\textbf{Topology of space at sub-quark level and masses of quarks and leptons}}
\author{%
\textbf{V.N. Yershov}\\                                          
\textit{University College London, Mullard Space Science Laboratory,}\\
\textit{Holmbury St.Mary, Dorking RH5 6NT, United Kingdom}\footnote{present address} \\
 \textit{Pulkovo Observatory, Pulkovskoye shosse 65-1, St.Petersburg 196140, Russia} \\
e-mail:\textit{vny@mssl.ucl.ac.uk}
}     

\date{}
%24.04.2003
\maketitle

\begin{abstract}
The global shape (topology) of the Universe is not derivable from 
General Relativity but should be determined by observations.  
Here we  propose a method for estimation of this shape using 
patterns of fundamental physical parameters,
for example the spectrum of fermion masses. We suppose that this pattern
might appear because of specific topology at sub-quark level.
We restrict ourselves to the analysis of a topological object
described by F.Klein in 1882 and show that 
its properties could give rise to structures reproducing 
three families of leptons and quarks.
\end{abstract}

\section{Introduction}

General relativity and the Big Bang theory assume that 
the universe is curved. But the character of this curvature 
(shape of the universe) cannot be deduced from
these theories without additional observations. Developments 
in theoretical and observational cosmic topology 
are progressing rapidly \cite{luminet}. Much observational 
effort has been directed towards the determination of the 
universe's curvature. Fewer observations have been aimed at   
topology. In the case of the universe with positive curvature 
there is a theoretical possibility to observe the same object
from different directions and to arrive to some conclusions
about the topology. But these observations are probably far
beyond our technical capabilities. Some non-trivial
features, such as multi-connectivity or the hypertoroidal
character of space, can probably be revealed by observing 
patterns in the distribution of galaxies and quasars \cite{roukema}, 
but the result will never be definitive. More accurate results 
one may probably obtain by high resolution measurements
of the cosmic microwave background radiation \cite{bond,levin}.

Here we would like to suggest a new method for determination
of the universe's shape. If some (or all) of the fundamental 
physical parameters were topology-dependent,
then one can try to find a proper topological model that
matches these parameters. Many physical parameters 
are known with high accuracy, and our method should be accurate
and more definitive than observations of
distant objects at the limits of our possibilities.

There is one set of parameters, which forms an enigmatic 
pattern probably related to the topology of space.
Since the discovery of quarks, it was found that there are only
twelve fundamental fermions grouped in three families 
(or generations) with  properties repeating from generation 
to generation. But masses of these particles \cite{properties} 
are distributed in a rather odd way (Table \ref{t:masses}). 

\begin{table}[htb]
\caption{Experimental rest masses of leptons \cite{properties} 
and quarks \cite{greene00} in units of the proton mass, $m_p$.}
\label{t:masses}
\begin{center}
\small
\begin{tabular} {cccccc} 
\hline \hline
\multicolumn{2}{c}{First generation} & \multicolumn{2}{c}{Second generation}
& \multicolumn{2}{c}{Third generation} \\ \hline
$\nu_e$ &  $\le 3\cdot 10^{-9}$  & $\nu_\mu$ & $\le 2\cdot 10^{-4}$ &
 $\nu_\tau$ & $\le 2\cdot 10^{-2}$ \\ %\hline
$e$ &  0.0005446170232(12)  & $\mu$ & 0.11260955173(34) & $\tau$ & 1.8939(3) \\
$u$ &  0.0047  & $c$ &  1.6 & $t$ & 189 \\ %\hline
$d$ & 0.0074   & $s$ &  0.16 & $b$ & 5.2 \\ \hline
\hline 
\end{tabular}
\end{center}
\end{table}

The Standard Model of particle physics uses some of  
these masses as its input parameters and absents from explaining
their origin. Although the experimental masses of quarks are not 
known with high accuracy, their distribution is wide and looks random.
This distribution can be considered as a good trial pattern for 
topological models. The masses of the charged leptons are known to high
accuracy, which gives one an opportunity to validate a model and 
estimate its accuracy.
Many theories have been proposed to explain this mass hierarchy
\cite{pati,salam} focusing basically on unification of all 
interactions \cite{georgi}, supersymmetric unification of 
fermions and bosons \cite{dimopoulos}, colour symmetries 
such as, e.g.,  technicolor \cite{weinberg}, or properties 
of one-dimensional objects (strings) \cite{green}. 
There are also theories investigating the possibility of randomness of
this pattern\cite{donoghue}. So far, none of these theories was able 
to approach the fermion mass distribution with a satisfactory 
accuracy. 

\section{Choise of topology}

Intuitively, A.Einstein and A.Friedmann imagined the 
universe as a 3-sphere of positive, negative
or zero curvature. But a 3-sphere, having its two 
hyper-interfaces (of course, if considered from the embedding
space), is not a good candidate for the shape
of the universe. By definition, the universe is self-contained,
and the existence of various interfaces might rise some doubts about 
this self-containedness. 
However, one can choose the well-known Klein bottle as an object 
corresponding to the universe's definition. 
This object, described by F.Klein in 1882 \cite{kastrup}, 
was a result of his work on the theory of invariants under group
projective  transformations.
Having a unique hyper-interface,  the universe with
 Klein bottle topology -- similar to the sphere -- 
can be of positive, negative or zero curvature.
The main feature of the Klein bottle hyper-surface is the unification of
its inner and outer interfaces. In our case, 
the unification might well occur at the sub-quark level, giving
 rise to substructures of the fermions.
Due to the clear pattern in properties of these 
particles (Table \ref{t:masses}) they cannot be considered 
as the elementary constituents of matter. It is logical to think 
that matter is structured further down to a simplest possible 
object (usually called ``preon'' referring to its primitive character). 
Here we shall consider the sub-quark unification area of the 
3-Klein-bottle as such a primitive particle.

\section{Primitive particles} 

The unification area of the 3-Klein-bottle can be considered 
as an area where space is turned ``inside-out''. One can attribute the known 
fundamental interactions to the geometrical properties of this inversion. 
For instance, distances $z$, measured from the preon's centre,
can be equivalently expressed in terms of their reciprocal values $z'=z^{-1}$,
if considered within the inverted manifestation of space 
\cite{greene00}. Thus, any potential, which is proportional 
to $1/z$, in the inverted manifestation of space will be proportional 
to $1/z'$ (that is, simply $z$). For instance, a Coulomb-like potential,
 $\phi_e \propto 1/z$,  in the inverted manifestation of space should 
manifest itself strong-likely, $\phi'_e \rightarrow \phi_s \propto z$, 
and vice versa. The distance $z_0$, where $\abs{z}=\abs{z'}$ and 
$\abs{\phi_s}=\abs{\phi_e}$, can characterize the scale, at which the 
preons stabilize forming structures. This use of classical potentials 
at sub-quark scales liberates us from the gauge anomaly problem. 
However, the problem of singularity (or energy divergences) precludes 
one from using Coulomb-like potentials near the centre of the source. 
Instead, one can use a potential self-cancelled at vicinity 
of $z=0$. Such a potential can be defined in many ways. 
As a simple example, one can take a potential derivable from a 
two-component field $F=F_0-F'_0$ with $F_0=s \exp(-1/z)$.
Here the signature $s$ indicates the charge of the trial particle
($s=\pm 1$), and the apostrophe denote the derivative with respect
to the radial coordinate $z$. In this example, two components 
of $F$ cancel each other out at a distance $z_0$ where $F_0=F'_0$, 
implying an equilibrium point between two interacting particles 
(the coupling constants are considered to be normalised to unity). 
Far from the source of the field the second component of $F$ mimics 
a Coulomb-like filed, whereas the first one extends to infinity 
being almost constant (similarly to the strong field). Here we are not 
going to discuss further details of the potential. What does matter
for our consideration is the symmetry of the potential resulting in the 
possibility for particles to couple irrespective of them
being like- or unlike-charged.

Consider a primitive particle (preon) possessing no properties 
except its charge and mass. For simplicity we shall use unit 
values for these quantities. We proceed on the Lorentz's 
premise that the particle mass is of purely electromagnetic origin.
Given the known three-polar (three-coloured) character of the 
strong interaction, we shall endue the preon's charge with colours,
labelling them as $red$, $green$, and $blue$. 
For convenient calculation of the signature $s$ one  
can use  a triplet of three-component column vectors
$\mathbf{\Pi}_i=\{\Pi_j\}_i$
 ($i,j=0,1,2$): 

\begin{equation*}
\Pi_{ij}=
 \begin{cases}
 +1, & i=j \\
 -1, & i\neq j 
 \end{cases}
\text{\hspace{0.6cm}}   \text{\hspace{0.2cm}} 
\overline{\mathbf{\Pi}}_i=-\mathbf{\Pi}_i.
%\label{eq:preonmatrix}
\end{equation*}
Then the signature $s_{ik}$ (normalised to unity) reads as 
\begin{equation}
s_{ik}=\pm \frac{\mathbf{\Pi}_i \cdot \mathbf{\Pi}_k}{\abs{\mathbf{\Pi}_i
\cdot  \mathbf{\Pi}_k}}
\label{eq:seforce}
\end{equation}
(positive and negative signs correspond respectively
to the strong and electric manifestations of the chromoelectric interaction).
Vectors $\mathbf{\Pi}_i$  also define the preons' unit charges 
$Z(\mathbf{\Pi}_i)=\sum_{j=0}^2{\Pi_{ij}}$
and unit masses
$m(\mathbf{\Pi}_i)=\abs{\sum_{j=0}^2{\Pi_{ij}}}$.
And the charge of a system composed of various preons or preon groups can be 
defined as
\begin{equation}
Z=\sum_{k=1}^N{\sum_{i=0}^2{\sum_{j=1}^n{\Pi_{ij}^k}}},
\label{eq:charge}
\end{equation} 
where  $N$ is the number of preon groups and $n$ is the number of preons in
the given group. We shall also define the mass $m$ and the  
reciprocal mass $m'$ of the system:

\begin{equation}
m= (1-\delta_{Z,0})
\sum_{k=1}^N\abs{{\sum_{i=0}^2{\sum_{j=1}^n{\Pi_{ij}^k}}}}, \hspace{0.7cm}
m'= (1-\delta_{Z,0})
\sum_{k=1}^N\abs{{\sum_{i=0}^2({\sum_{j=1}^n{\Pi_{ij}^k})^{-1}}}},
 \label{eq:mass}
\end{equation} 
where $\delta_{Z,0}$ is the Kronecker delta-function, implying
the Lorentz's conjecture: if two oppositely charged particles combine
(say $red$ and $antigreen$), not only their charges but also
their masses are neutralised. 
Eq.(\ref{eq:charge}) and  (\ref{eq:mass}) are approximate because
they do not take into account residual polarisation existing at the 
vicinity of any dipole. 
The complete cancellation of charges (hence -- masses) wight be possible 
only if the centres of the interacting
particles were coinciding, which is not exactly our case (the preons
in our model are separated at least by the distance $z_0$). 
But for the sake of simplicity, we shall neglect 
small residual masses of the  neutral preon structures.  

\section{Combinations of the primitive particles}

The signature (\ref{eq:seforce}) necessarily implies a group of 
combinative preon structures. The simplest structure is a charged  
preon doublet $(\mathbf{\Pi}_i,\mathbf{\Pi}_k)$:
\begin{equation*}
\varrho^\pm_{ik}= \mathbf{\Pi}_i+\mathbf{\Pi}_k, \text{\hspace{0.6cm}} i,k=0,1,2
\end{equation*}
(six possible combinations for $\mathbf{\Pi}$ and six others for
$\overline{\mathbf{\Pi}}$). A neutral preon doublet 
$(\mathbf{\Pi}_i,\overline{\mathbf{\Pi}}_k)$:
\begin{equation*}
g^0_{ik}= \mathbf{\Pi}_i+\overline{\mathbf{\Pi}}_k 
\end{equation*}
(nine combinations)  is also possible.
The preon doublets will be deficient in one or two colours.
According to (\ref{eq:charge}) and 
(\ref{eq:mass}), 

\begin{equation*}
Z(\mathbf{\Pi}_i,\mathbf{\Pi}_k)=\pm 2, \hspace{0.4cm}
m(\mathbf{\Pi}_i,\mathbf{\Pi}_k)=2, \hspace{0.4cm}
m'(\mathbf{\Pi}_i,\mathbf{\Pi}_k)=\infty,
\end{equation*}

\begin{equation*}
Z(\mathbf{\Pi}_i,\overline{\mathbf{\Pi}}_k)=0, \hspace{0.4cm}
m(\mathbf{\Pi}_i,\overline{\mathbf{\Pi}}_k)=0, \hspace{0.4cm}
m'(\mathbf{\Pi}_i,\overline{\mathbf{\Pi}}_k)=\infty.
\end{equation*}

Then, if an additional charged preon is added to the neutral
doublet, the mass and the charge of the system are restored, according
to (\ref{eq:mass}): 

\begin{equation}
m(\mathbf{\Pi}_r,\overline{\mathbf{\Pi}}_g,\mathbf{\Pi}_b)=1
\text{\hspace{1.0cm} but still \hspace{1.0cm}} 
m'(\mathbf{\Pi}_r,\overline{\mathbf{\Pi}}_g,\mathbf{\Pi}_b)=\infty.
\label{eq:mgluon}
\end{equation}

The charged doublets $2\Pi$ and  $2\overline{\Pi}$
 will not be free for long because their colour potentials 
($\phi_s \propto z$) extend  to infinite
distances. Any distant preon of the same charge but with a
complementary colour will be attracted to
the pair. In this way,  $3\Pi$ and  $3\overline{\Pi}$ {\sf Y}-shaped  
particles will be formed. The mass of the {\sf Y}-particle
corresponds to the three unit preon's masses, and its charge (positive
or negative) is of the same magnitude. Its colour will be 
complete, but locally, the $rgb$ colours  of its three preons 
will be distributed in a plane forming a closed loop. Thus,
a part of the strong field is ring-closed in this plane,
whereas another is extended to infinity (over the ring's poles).

\begin{figure}[htb]
\IfFileExists{graphicx.sty}{
\centerline{  \includegraphics[width=0.6\textwidth]{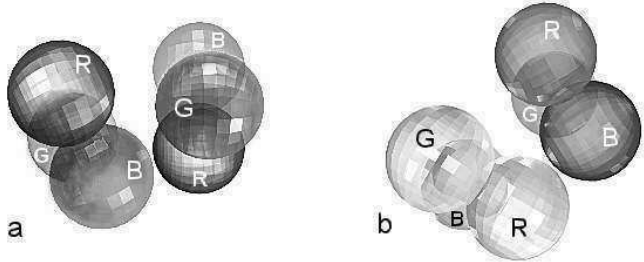}}
}{
  Sorry, package \texttt{graphicx} not present.
}
\caption{({\bf a}) Two combined like-charged
{\sf Y}-particles turned through $180^\circ$ with respect 
to one another; and ({\bf b}) two coupled unlike-charged {\sf Y}-particles
turned through $120^\circ$ with respect to one another.}
\label{fig:twoy}
\end{figure}

{\sf Y}-particles cannot be free because their strong potentials 
are only partially closed within loops. 
Thus, distant like-charged {\sf Y}-particles will combine  
and form {\sf YY}  structures
(Fig.\ref{fig:twoy}{\bf a}).  The planes of the paired triplets 
are parallel to each other. The second 
particle is turned through $180^\circ $ in respect with the first one.
This is the only possible mutual orientation of two
combined like-charged {\sf Y}-particles if there are no other 
particles at the vicinity of the pair.  
The colour pattern of this structure can be written as, e.g.,   

\begin{equation}
\begin{vmatrix} r_1 & & & b_2 \\  & g_1 &
g_2 &   \\ b_1 & & & r_2 \end{vmatrix}.
\label{eq:antiparallel}
\end{equation}

Similarly, in the absence of other particles two unlike-charged
{\sf Y}-particles will couple rotated mutually through $180^\circ $,
forming a neutral structure. Otherwise they
combine by turning clockwise through $120^\circ$
with respect to one another ($\uparrow$), Fig.\ref{fig:twoy}{\bf b},
or anticlockwise ($\downarrow$), with two 
corresponding colour patterns:

\begin{equation}
\begin{vmatrix} \overline{g}_1 & & r_2 &  \\  & \overline{b}_1 &
& g_2   \\ \overline{r}_1 & & b_2 &  \end{vmatrix}
 \text{\hspace{0.5cm} or \hspace{0.5cm} }   
\begin{vmatrix} \overline{g}_1 & & b_2 &  \\  & \overline{b}_1 &
& r_2   \\ \overline{r}_1 & &
g_2 &  \end{vmatrix}. \label{eq:parallel}
\end{equation}

The three-colour completeness of {\sf Y} permits up to three of them
to combine if all of them are like-charged. These three {\sf Y}
will be joined in a closed loop with the following distribution of colours:

\begin{equation}
\begin{vmatrix}  & r_1 &  & b_2 & & g_3 \\
 b_1 &  & g_2  &  & r_3 &  \\
  & g_1 &  & r_2 & & b_3  \end{vmatrix}
 \text{\hspace{0.5cm} or \hspace{0.5cm} } 
\begin{vmatrix}  & r_1 &  & g_2 & & b_3 \\
 b_1 &  & r_2  &  & g_3 &  \\
  & g_1 &  & b_2 & & r_3  \end{vmatrix},
\label{eq:threey}
\end{equation}
corresponding to the clockwise and anticlockwise mutual orientation
of the components.

\begin{figure}[htb]
\IfFileExists{graphicx.sty}{
\centerline{  \includegraphics[width=0.6\textwidth]{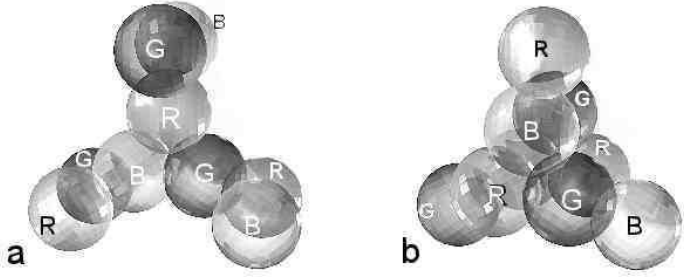}}
}{
  Sorry, package \texttt{graphicx} not present.
}
\caption{ Three like-charged {\sf Y}-particles joined with their vertices 
directed ({\bf a}) towards the centre of the structure (left-handed state)
and ({\bf b}) away from the centre (right-handed state).  }
\label{fig:electron}
\end{figure}
These structures are shown in Fig.\ref{fig:electron}{\bf a}, where the
vertices of their {\sf Y}-components are directed towards their common centre.
A state with the vertices of {\sf Y}  directed away from their
common centre is also possible (Fig.\ref{fig:electron}{\bf b}).
The latter can be obtained from the former  
by mirror-reflection of all its components about the 
circular axis of symmetry. 
We shall refer to these two states as the left- and right-handed ones
(3{\sf Y}$_L$ and 3{\sf Y}$_R$).
The mass of the 3{\sf Y}-particle is the sum of masses of its
nine preons (9 units). Similarly, its charge  is 9 preon
charge units.
Connecting the like-coloured preons of this structure, it is 
seen that the spatial distribution of any particular colour 
appears as a helical trajectory (current) twisted along the loop.
These colour-charge  currents (supposedly related to the gyromagnetic
properties of the structure) might be twisted clock- or 
anticlockwise with respect to the loop's circular axis.

Couples of the unlike-charged {\sf Y} can form chains
{\sf Y}$\overline{\sf Y}$ - {$\overline{\sf Y}$}{\sf Y} -
 {\sf Y}$\overline{\sf Y}  \ldots$
 with the following colour patterns:

\begin{multline}
\left|
\begin{smallmatrix} & r_1 & &  \overline{b}_2 \\ b_1 &
& \overline{g}_2 &   \\ & g_1 & & \overline{r}_2
\end{smallmatrix}
\right| +
\left|
\begin{smallmatrix} \overline{r}_3 & & b_4 &  \\  &
\overline{g}_3 &  & r_4  \\ \overline{b}_3 & & g_4 &
\end{smallmatrix}
\right| +
\left|
\begin{smallmatrix}  & g_5 &  & \overline{r}_6 \\ r_5
&  & \overline{b}_6 &  \\  & b_5 &  & \overline{g}_6
\end{smallmatrix}
\right| +
\left|
\begin{smallmatrix} \overline{g}_7 & & r_8 &  \\  &
\overline{b}_7 &  & g_8  \\ \overline{r}_7 & & b_8 &
\end{smallmatrix}
\right| +
\left|
\begin{smallmatrix}  & b_9 &  & \overline{r}_{10} \\
g_9 &  & \overline{r}_{10} &   \\ & r_9 &  & \overline{b}_{10}
\end{smallmatrix} \right| +
\left|
\begin{smallmatrix} \overline{b}_{11} & & g_{12} &  \\
& \overline{r}_{11} & & b_{12}  \\ \overline{g}_{11} & & r_{12} &
\end{smallmatrix} \right| + \dots
\label{eq:nuright}
\end{multline}

\begin{multline}
\left|
\begin{smallmatrix} & r_1 & &  \overline{g}_2 \\ b_1 &
& \overline{r}_2 &   \\ & g_1 & & \overline{b}_2
\end{smallmatrix}
\right| +
\left|
\begin{smallmatrix} \overline{b}_3 & & r_4 &  \\  &
\overline{r}_3 &  & g_4  \\ \overline{g}_3 & & b_4 &
\end{smallmatrix}
\right| +
\left|
\begin{smallmatrix}  & b_5 &  & \overline{r}_6 \\ g_5
&  & \overline{b}_6 &  \\  & r_5 &  & \overline{g}_6
\end{smallmatrix}
\right| +
\left|
\begin{smallmatrix} \overline{g}_7 & & b_8 &  \\  &
\overline{b}_7 &  & r_8  \\ \overline{r}_7 & & g_8 &
\end{smallmatrix}
\right| +
\left|
\begin{smallmatrix}  & g_9 &  & \overline{b}_{10} \\
r_9 &  & \overline{g}_{10} &   \\ & b_9 &  & \overline{r}_{10}
\end{smallmatrix} \right| +
\left|
\begin{smallmatrix} \overline{r}_{11} & & g_{12} &  \\
& \overline{g}_{11} & & b_{12}  \\ \overline{b}_{11} & & r_{12} &
\end{smallmatrix} \right| + \dots
\label{eq:nuleft}
\end{multline}
corresponding to two ($\uparrow$ and $\downarrow$) possible 
states (\ref{eq:parallel}).
The colour patterns repeat after each  six consecutive
${\sf Y} \overline{\sf Y}$ groups, forming   
12{\sf Y}-period chains.  The 12-th chain element
is compatible with the first one, which makes chains to 
close in loops (with the 12{\sf Y}-ring being the minimal-length
loop).

\begin{figure}[htb]
\IfFileExists{graphicx.sty}{
\centerline{  \includegraphics[width=0.6\textwidth]{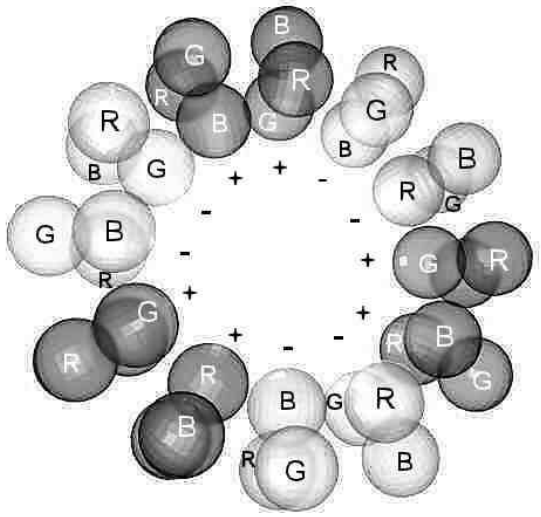}}
}{
  Sorry, package \texttt{graphicx} not present.
}
\caption{12{\sf Y}-structure.
 Mirror-reflection of all the components in   
 the loop about the loop's circular axis translates
 the structure into itself because of the equal number of
 {\sf Y} directed with their vertices towards and outwards the 
 centre of the structure.}
\label{fig:twonu}
\end{figure}

The pattern (\ref{eq:nuright}) is  visualised in
Fig.\ref{fig:twonu}, where brighter colours are assigned 
to the negatively charged preons.
These ring-closed chains consisting of twelve {\sf Y}-particles
are neutral and almost massless, according  
to (\ref{eq:charge}) and (\ref{eq:mass}).
The spatial distribution of any particular colour
appears as a clockwise ($\uparrow$) or anticlockwise
($\downarrow$) helix, each being a complete twist 
around the closed loop axis.

It is interesting to note that the mirror-reflection of all the 
 components of 12{\sf Y} about its circular axis translates 
the structure into itself. Thus,  12{\sf Y}$_L$ and  12{\sf Y}$_R$ 
are topologically indistinguishable because the number of
their {\sf Y}-components directed with their vertices 
inwards the loop coincides with the number of those 
directed outwards.

The 12{\sf Y}-particle, consisting of 36 preons, is massless
unless coupled to a charged particle, say {\sf Y} or 3{\sf Y},
which restores its mass. 
 For instance, according to (\ref{eq:mass}),
 the mass of {\sf Y}$^*$={\sf Y}+12{\sf Y} 
 is 39 preon mass units (3+36).
The (3{\sf Y}+12{\sf Y})-particle is of 45 mass units (9+36), etc.
Coupling {\sf Y} with 12{\sf Y} and 3{\sf Y} with 12{\sf Y} 
is possible because of attractive forces arising due to 
the particles' local patterns of colours and charges
(analogous to the van der Waals forces between molecules).
The strength and the sign of this force depends on the compatibility
of colour patterns (helices) of the interacting particles.
The colour pattern of {\sf 12Y$_\uparrow$}
does not match that of {\sf 3Y$_\downarrow$}.
Only {\sf 3Y$_\uparrow$} and {\sf 12Y$_\uparrow$} or 
{\sf 3Y$_\downarrow$} and {\sf 12Y$_\downarrow$} can combine.
 Unlike this case, if {\sf 3Y} combines with another
 {\sf 3Y}, or {\sf 12Y} with another {\sf 12Y}, their helices  
should be opposite.   

By their properties, 3{\sf Y} and 12{\sf Y} can be readily
associated with the electron and its neutrino (leptons of the first
generation).  The charge of 3{\sf Y}, divided by 9, gives us the
conventional unit charge of the electron. 
Then charges of {\sf Y}$^*$ or 2{\sf Y}$^*$-particles
correspond to the fractional charges of 1/3 and 2/3 
(hinting at possible quark's constituents).
The {\sf Y}$^*$-structure cannot be free because of the 
 strong potential of its central {\sf Y}-component. It will further
combine with other {\sf Y}$^*$. If two {\sf Y}$^*$-particles 
have likewise helix patterns, they will couple via an intermediate 
 12{\sf Y}-particle with an opposite-helical pattern. 
The {\sf Y}$^*_\uparrow$ {\sf 12Y}$_\downarrow$ {\sf Y}$^*_\uparrow $-link
can be identified with the $up$ quark. It will be charged (with a 
charge +3+3=+6 units) and having a mass (39+39=78 units).

\begin{table}[htb]
\caption{Simple preon structures.  A similar table can be constructed
for antipreons (with opposite charges in the third column). 
}
\label{t:preonsummary}
\begin{center}
%\small
\begin{tabular}
{ccrc} \hline \hline

Structure     & Constituents       & Charge         & Mass            \\
              & of the             & (in preon     & (in preon      \\
              & structure          & charge units) & mass units)  \\ \hline
\multicolumn{4}{c} {Primitive particle (preon)} \\ %\hline
$\Pi$  & $1\Pi$                &  $-1$       &   1           \\ 
\hline
\multicolumn{4}{c} {Structures consisting of single preons} \\ %\hline
$\varrho^-$       & $2\Pi$       &  $-2 $       &   2           \\ 
$g^0$         & $1\Pi+1\overline{\Pi}$ & $-1+1=0$  &   0           \\
{\sf Y} & $3\Pi$             & $-3 $        &   3           \\  
\hline
\multicolumn{4}{c}{Second order structures consisting of preon triplets ({\sf Y}-particles)} \\
%\hline 
$\delta^-$  & 2{\sf Y}         & $-6 $       &   6            \\ 
$\gamma^0$    & $1{\sf Y}+1\overline{\sf Y}$ & $-3+3=0$   &   0             \\ 
$e^-$       & 3{\sf Y}         & $-9 $       &   9             \\ \hline
\multicolumn{4}{c}{Structures consisting of the second order substructures}  \\
%\hline
$2e^-$      & $3{\sf Y}+3{\sf Y}$ & $-18$       &  18 \\
$2e^0$        & $3{\sf Y}+3\overline{\sf Y}$    & $-9+9=0$              &  0 \\
$\nu^0$       & 12{\sf Y}    & $6\times (-3+3)=0$              &  0  \\
{\sf Y}$^*$ & $\nu$ \ \   + \ \  {\sf Y}   & $0-3=-3$   &  39 \\
$W^-$   & $\nu$ \ \ + \ \ $e$            & $0-9=-9$        &  45 \\
$u^+$       & {\sf Y}$^*$ \ \ $\nu$ \ \ {\sf Y}$^*$  & $+3+0+3=+6$        &  78 \\
$\nu^0_\mu$   & {\sf Y}$^*$ \ \ $\nu$ \ \ $\overline{{\sf Y}}^*$      &  $-3+0+3=0$             &  0 \\
$d^-$       & $u$ \ \ + \ \ $W$  & $+6-9=-3$ & 123 \\
$\mu^-$     & \ $\nu_{\mu}$ \  + \ \ $W$  & $0-9=-9$ & $(48,39)^*$ \\
\multicolumn{4}{c}{and so on ...}  \\     
\hline \hline
\end{tabular}
\end{center}
{\footnotesize \hspace{3cm}*)two-component system }
\end{table}

Being positively charged, the $up$ quark can couple to a
negative particle, such as $3\overline{\sf Y} 12{\sf Y}^0$ 
(with its 45 mass units and its -9 charge units).
The resulting mass of $3\overline{\sf Y}12{\sf Y}^0$ + 
{\sf Y}$^*$ 12{\sf Y}$^0$ {\sf Y}$^*$
(the $down$ quark) is $123= 45+78$, and its charge is
$-3=-9+6$. Simple preon structures are summarised in
Table \ref{t:preonsummary}.

\section{The second and third generations of particles}

It is natural to suppose  that fermions of the second and third
generations should be composed of simpler structures belonging to the 
first generation.
For instance, the muon neutrino  (a neutral particle) can be formed
of unlike-charged {\sf Y}$^*$={\sf Y$ \nu_e$} and
$\overline{\sf Y}^*=\overline{\sf Y} \nu_e$:

\begin{equation}
\nu_\mu= {\sf Y} 12{\sf Y}^0_\downarrow 
 \ 12{\sf Y}^0_\uparrow \  12{\sf Y}^0_\downarrow  \overline{\sf Y} 
={\sf Y}^*\nu_e\overline{\sf Y}^*, 
 \label{eq:numu}
\end{equation}
and the muon could be structured naturally as

\begin{equation}
\mu= (12{\sf Y}^0+3\overline{\sf Y}{\sf Y})
(12{\sf Y}^0_\downarrow 12{\sf
Y}^0_\uparrow  12{\sf Y}^0_\downarrow \overline{\sf Y})
=\overline{\nu}_e e^-\nu_\mu ,
\label{eq:muon}
\end{equation}
and so on. 
These structures, depending on their complexity, 
can be rigid or non-rigid. In our model, the fermions 
belonging to the second and third generations are considered as clusters,
rather than rigid structures 
(in (\ref{eq:muon}) the clustered components are enclosed in parentheses).
Their masses depend on the sum $M=\sum_k{m_k}$ of the masses of their 
components, $m_k$, and the sum of inverted reciprocal masses of the 
components, $1/m'_k$: $\frac{1}{M'}=\sum_k{\frac{1}{m'_k}}$. 
The combined mass can be computed as 

\begin{equation}
m(m_1, m_2, \ldots , m_N)=M M'=\frac{m_1+m_2+\dots+m_N}
{1/m'_1+1/m'_2+\dots+1/m'_N} .
\label{eq:mtotal}
\end{equation}

Using this formula for fermions and comparing their computed masses with 
the experimental data, one can see that, 
for the second and third generations, the masses are reproduced 
with a systematic error of about 0.5\% (we do not
present these results here). The systematic differences must be 
attributed to the already mentioned simplifications in (\ref{eq:mass}), 
as well as to the neglect of relativistic effects and dynamics of 
colour-charge currents responsible for the gyromagnetic properties of 
the structures. However, it is seen that the systematic trend 
depends on the number of preons in the clustered components 
containing 3{\sf Y} and can be readily taken into account by small 
corrections to the masses of the clustered components as following:

\begin{equation}
m_i^{c}=m_i+\varkappa  \eta,
\label{eq:corrmass}
\end{equation}
where $m_i$ is the original mass of the $i$-th component (in  
units of the preon's mass), $m_i^{c}$ is the corrected mass of the component,
and $\eta$ is the correction factor:

\begin{equation}
\eta=\frac{1}{54\pi^2}\sqrt{\frac{m_e}{m_e^{c}}}.
\label{eq:mcorr}
\end{equation}
 $m_e^{c}$ is the corrected electron charge
calculated from the recursive expression

\begin{equation*}
m_e^{c}=m_e-m_e^{c}\eta,
\end{equation*}
 $m_e^{c}$ is the corrected mass of the electron,
calculated recursively by using (\ref{eq:corrmass});
$m_e$ is the original electron mass, expressed in 
preon units ($m_e=9$); and 
$\varkappa$ is the preon factor for the  
given clustered component: 

\begin{equation*}
\varkappa= \left\lbrace
 \begin{array}{ll}
 \frac{N^+}{N^-} m_e, & {\rm for \hspace{0.2cm} the \hspace{0.2cm} 
positively \hspace{0.2cm} charged \hspace{0.2cm} components, \hspace{0.2cm}
 if \hspace{0.3cm}}  Z_{system}>0 \\
 -N^-, & {\rm for \hspace{0.2cm} the \hspace{0.2cm} negatively \hspace{0.2cm}
charged \hspace{0.2cm} components, \hspace{0.2cm}
 if \hspace{0.3cm}}  Z_{system}<0.  
 \end{array}
 \right.
\end{equation*}
Here $N^+$ and $N^-$ are respectively the preon numbers in the 
positively and negatively charged clustered components.
 For $m_e=9$ and $\varkappa_e=-9$ the correction 
factor (\ref{eq:mcorr}) is  $\eta=0.001878079$ ($m_e^{c}=8.98313$).
The constant in (\ref{eq:mcorr}) is tuned for the best fit of the 
systematic trend in (\ref{eq:mtotal}).
%%%%

The fermion masses obtained with the use of (\ref{eq:mtotal}) and 
(\ref{eq:corrmass}) are summarised in Table \ref{t:massresult}. 
As an  example, we can compute the muon's mass. The masses of the 
muon's  components, according to its structure (\ref{eq:muon}), are: 
$m_1=m'_1=48$, $m_2=m'_2=39$, $\varkappa=-123$, 
$m_2^{c}=38.768996$ (all in preon mass units). And the muon's mass is

\begin{equation*}
m_\mu (m_1,m_2^{c}) = \frac{m_1+m_2^{c}}{\frac{1}{m'_1}+
\frac{1}{m_2^{'c}}}=1860.91182
 {\hspace{0.2cm} {\rm (preon \hspace{0.2cm} mass \hspace{0.2cm} units)}.}
\end{equation*}

For the $\tau$-lepton:
 $m_1=m'_1=156$, $m_2=m'_2=201$,
$\varkappa=-201$, $m_2^{c}=200.6227$,   

\begin{equation*}
m_\tau (m_1,m_2^{c})=31297.1416 {\hspace{0.2cm} {\rm (preon 
\hspace{0.2cm}  mass \hspace{0.2cm} units)}.}
\end{equation*}

For the proton, positively charged particle consisting of two $u-$,
one $d-$quarks and a cloud of gluons $g$, masses of its components 
are $m_u=m'_u=78$,  $N^+=78$,  $m_d=m'_d=123$, $N^-=123$ 
($m_u^{c}=78.0107132$). As for the gluons, only those of them
should be taken into account, which, at any given moment of time,  
are coupled to the quarks' preons because otherwise these gluons 
are massless.
The total number of the coupled gluons, in accordance with the proton's
structure ($2u+d$), is $N_g=2N^++N^-=279$.
The mass of each gluon coupled to a preon, according to (\ref{eq:mgluon}),
 is $m_g=1$, $m'_g=\infty$. Then the resulting proton mass is 

\begin{equation}
m_p  = \frac{2m_u^{c}+m_d+N_gm_g}{2\frac{1}{m_u^{'c}}+
\frac{1}{m'_d}+N_g\frac{1}{m'_g}}=
16525.3588 {\hspace{0.2cm} {\rm (preon \hspace{0.2cm} mass \hspace{0.2cm} units)}.}
\label{eq:pmass}
\end{equation}

Using  (\ref{eq:pmass}), one can convert $m_e$, $m_\mu$, $m_\tau$ and 
masses of other particles from preon mass units into the proton
mass units, $m_p$. These values are given  in the fourth column 
of Table \ref{t:massresult}. 
The experimental masses of the particles 
(also expressed in units of $m_p$) are listed in the last
column for comparison.

\begin{table}[htb]
\caption{Predicted and experimental rest masses of quarks and leptons.
Values given in the third column can be  converted into the proton
mass units dividing them by $m_p=16525.3588$. Experimental masses in the 
last column are taken from Ref.\cite{properties} for leptons and 
from Ref.\cite{greene00} for quarks.}
\label{t:massresult}
\begin{center}
\small
\begin{tabular}
{lcllll} \hline \hline
\multicolumn{2}{c}{Particle and}  & {Number of active}  &  Computed    & Computed  & Experimental \\
\multicolumn{2}{c}{its structure} & {preons (composite}  & masses (preon & masses   & masses    \\
\multicolumn{2}{c}{ (components)} & mass) & mass units)   & (in $m_p$) &  (in $m_p$) \\ \hline
\multicolumn{6}{c} {First generation} \\ 
%\hline
$\nu_e$ & 12{\sf Y}$^0$            & 0   & $\approx 0$    & $\approx 0$       & $<3\cdot 10^{-9}$ \\
$e^-$ & $3\overline{{\sf Y}}$               & 9    & 9    & 0.0005446175 & 0.0005446170 \\ 
$u$ & {\sf Y}$^*\nu_e${\sf Y}$^*$ & 78   & 78   & 0.004720019     & 0.0047 \\
$d$ & $u$ $ \ \nu_e e^-$                 & 123  & 123  & 0.007443106     & 0.0074 \\
\hline
\multicolumn{6}{c}{Second generation} \\
%\hline
$\nu_\mu $ & {\sf Y}$^*$ \ \ $\nu_e$  \ \  $\overline{{\sf Y}}^*$ & 0  & $\approx 0$ & $\approx 0$ & $<2\cdot 10^{-4}$  \\
$\mu^-$ & \ \ $\nu_\mu$ \ \ + \   $\nu_e e^-$    & $m(48, 39)$   & 1860.9118 & 0.11260946 & 0.11260951 \\
$c$ & {\sf Y}$^{**}$ \ \ + \ \ {\sf Y}$^{**}$ & $m(165, 165)$ & 27122.89     & 1.641289  & 1.6 \\
$s$ &  \ $c$ \ \ \ + \ \ $e^-$ & $m(165, 165, 9)$ & 2745.37 & 0.1661307 & 0.16 \\
\hline
\multicolumn{6}{c}{Third generation} \\
%\hline
$\nu_\tau$ & $u$ \ \ \  $\nu_e$ \ \ \  $\overline{u}$ & 0  & $\approx 0$ & $\approx 0$ & $<2\cdot 10^{-2}$ \\
$\tau^-$ & \ \ $\nu_\tau$ \ \  + \    $\nu_\mu \mu^-$ & $m(156, 201)$ & 31297.11 & 1.893884 & 1.8939 \\
$t$ & {\sf Y}$^{***}$ \  + \ \ {\sf Y}$^{***}$ & $m(1767, 1767)$ & 3122289 & 188.9392 & 189 \\
$b$ & \  $t$ \ \ + \ \ $\mu^-$ & $m(1767, 1767, 48, 39)$ & 75813.33 & 4.587696 & 5.2  \\
\hline \hline
\end{tabular}
\end{center}
\end{table}

In this Table, {\sf Y}$^*$, {\sf Y}$^{**}$ and {\sf Y}$^{***}$ stand for the structures
with triplets {\sf Y} coupled to ring-like particles $\nu$ of increasing complexity. 
For instance, the electron-neutrino, $\nu_e=12{\sf Y}$, gives rise 
to a particle ${\sf Y}^*=\nu_e{\sf Y}$. 
Ring structures similar to that of the the electron neutrino,
may be considered as ``heavy neutrinos'',
$\nu_h =12{\sf Y}^*$. They
can further form ``ultra-heavy'' neutrinos
$\nu_{uh} =3(\overline{\sf Y}^* \nu_h u)e^-$, and so on, with the number
of preons increasing with the complexity of the structure. 
The components {\sf Y}$^{**}$ and {\sf Y}$^{***}$ of  
$c$ and $t$ may have the following structures: 
{\sf Y}$^{**}=u\nu_e u\nu_e e^-$, consisting of 165 preons, 
and {\sf Y}$^{***}= \nu_{uh}${\sf Y} consisting of 1767 preons.

Table \ref{t:massresult}
illustrate family-to-family similarities between the  particle
structures. For example, in each family, the $down$-like quark appears
as a combination of the $up$-like quark, with a charged lepton belonging 
to the lighter family. Thus, according to this scheme,
 the {\it strange}-quark is composed of the {\it charm}-quark
 and the electron:
$s=c+e^- $
and has a  mass,
$m_s(165,165,9)=2745.37$, resulting from $m_c$(165,165)=27122.89
and $m_e=9$. Similarly,  the mass of the {\it bottom}-quark,
$b=t+\mu^-$, 
is a combination
of $m_t$(1767,1767)=3122289 with $m_\mu$(48,39)=1860.9118, resulting in 
$m_b(1767,1767,48,39)=75813.33$. The parentheses notation here 
corresponds to  the abbreviated writing of (\ref{eq:mtotal}). 
 Each charged lepton is a combination of
the neutrino from the same family with the neutrino and the charged
lepton from the lighter family:
$\mu^-=\nu_\mu+\overline{\nu}_ee^-$
$\tau^-=\nu_\tau+\overline{\nu}_\mu\mu^-$.

\section{Conclusions}

The discussed model, using the Klein-bottle topology of the universe, 
has reproduced the pattern of the fermion 
masses without using experimental input parameters.   
The computed masses of the charged leptons agree with experiment
to an accuracy of about $10^{-6}$. The predicted masses of the quarks
are also in good agreement with experiment. It is quite improbable
that nine derived quantities could just spuriously agree with 
the seemingly random distributed experimental fermion masses. 
Thus, we have to conclude that the universe possesses 
the Klein-bottle-like topology.

\end{document}